\documentstyle[12pt,fps,epsfig]{article}
\newcommand{\Z}{Z \!\!\! Z}

\newcommand{\vp}{\varphi}

\newcommand{\tila}{\tilde \lambda}
\makeatletter
\@addtoreset{equation}{section}
\makeatother

\begin{document}
\vspace*{1cm}
\begin{center}
{\Large Perfect Lattice Perturbation Theory: \\
\ \\
A Study of the Anharmonic Oscillator}

\vspace*{1cm}

W. Bietenholz$^{\rm a}$ and T. Struckmann$^{\rm b}$
\vspace*{7mm}

$^{\rm a}$ HLRZ c/o Forschungszentrum J\"{u}lich \\
52425 J\"{u}lich, Germany \\

\vspace*{7mm}

$^{\rm b}$ Physics Department \\
University of Wuppertal \\
D-42097 Wuppertal, Germany

\vspace*{1cm}

Preprint HLRZ 1997-67, WUB 97-35

\end{center}

\vspace*{1cm}

As an application of perfect lattice perturbation theory,
we construct an $O(\lambda)$ perfect lattice action
for the anharmonic oscillator analytically in
momentum space. In coordinate space we obtain
a set of 2-spin and 4-spin couplings $\propto \lambda$,
which we evaluate for various masses. These couplings
never involve variables separated by more than two
lattice spacings.

The $O(\lambda)$ perfect action is simulated and compared
to the standard action. We discuss the improvement for
the first two energy gaps $\Delta E_{1}$, $\Delta E_{2}$
and for the scaling quantity  $\Delta E_{2}/\Delta E_{1}$
in different regimes of the interaction parameter, and of the
correlation length.


\section{Introduction}

The only non-perturbative access to complicated 4d quantum field 
theories, such as QCD, which proved successful, are Monte Carlo
simulations on the lattice. They necessarily take place at a
finite lattice spacing $a$ and in a finite size $L$. In order
to reveal information about continuum physics in an infinite
volume, we have to require $a << \xi << L$, where $\xi$
is the correlation length. In particular the finiteness
of $\xi /a$ causes serious systematic errors in practical simulations.
It is now very fashionable to fight such artifacts by using
``improved lattice actions'' \cite{LAT97}. These are discretizations
of the continuum action, which are supposed to display
the correct scaling behavior down to a much shorter correlation
length in lattice units, than it is the case for the standard 
lattice action.

In the literature, there are mainly two strategies to construct
improved lattice actions, in particular for QCD.
The first one is called Symanzik's program \cite{Sym}. One tries
to eliminate the lattice spacing artifacts order by
order in $a$ -- similar to the Runge and Kutta procedure for
the numerical solution of ordinary differential equations.
This is achieved by adding irrelevant operators.
On the classical level, and in the framework of on-shell
improvement \cite{OnShell},
the standard Wilson action for QCD 
could be improved to $O(a)$ analytically
by adding the so-called clover term \cite{SW}. 
On the quantum level, the coefficient of this term gets 
renormalized, and the quantum correction was first
estimated numerically by a mean field approach \cite{Tad}.
The complete $O(a)$ improvement was finally determined
by the ALPHA collaboration based on extensive simulations
\cite{alpha}. However, it seems hardly feasible to carry
on this program beyond $O(a)$.

The alternative method uses renormalization group concepts
to construct quasi-perfect actions. These are approximations
to perfect actions, i.e. to actions which are completely free 
of cutoff artifacts \cite{WilKog}.
As a fundamental difference from Symanzik's program, this method
is non-perturbative with respect to $a$.
As a first step, this program can be realized perturbatively
(in the interaction),
which yields analytic expressions for the perfect quark-gluon 
and 3-gluon vertex functions \cite{QuaGlu,StL,KODiss}.
\footnote{Perturbatively perfect actions have also be studied
for the Schwinger model, \cite{Schwing,Farc}.} 
Thus one eliminates all artifacts of $O(a^{n})$ and $O(ga^{n})$,
such that the remaining artifacts are of $O(g^{2}a)$ and beyond
($g$ is the gauge coupling).
This is opposed to the action of Ref. \cite{alpha}, which is
free of artifacts in $O(g^{n}a)$, but plagued for instance by
systematic errors in $O(a^{2})$.

An extension of this program is the construction of ``classically
perfect actions'' \cite{Has}. This approach, which is designed
particularly for asymptoti-cally free models, is non-perturbative
also with respect to the coupling $g$.
Using a multigrid procedure,
one identifies the fixed point action of an renormalization group
transformation. This can be
done solely by minimization -- the functional integral reduces
to a classical field theory problem -- and the fixed point action
then serves as an approximatively perfect (``classically perfect'')
action at finite correlation length too. 
In a sequence of toy models, it turned out that classically perfect 
actions are excellent approximations to (quantum) perfect actions,
in the sense that they drastically suppress lattice spacing artifacts.
The improvement achieved in this way goes far
beyond first order Symanzik improvement. This has been observed for
the 2d O(3) and CP(3) model \cite{Has,Ruedi}, the Schwinger model 
\cite{Lang} and the 1d XY model \cite{rotor}.
In principle that program can be extended also beyond classical
perfection, if one performs e.g. one real space MCRG step at finite
correlation length, starting from a classically perfect action.

The construction, which is non-perturbative in $a$ and in $g$,
is presumably the climax of the improvement program.
However, in perfect and also in classically perfect actions the 
couplings tend to involve infinite distances, and we can at best
achieve locality in the sense of their exponential decay.
For practical purposes a truncation is needed, which does
some harm to the quality of the improvement. This is the main reason
why the second, more sophisticated, improvement program
could not be applied yet in a satisfactory way to QCD.\\

Here we focus on the perturbatively perfect action.
It has potential applications with two respects:
it can either be used directly, or as a starting point
of the non-perturbative multigrid improvement \cite{StL}.
A direct application of a truncated perfect quark-gluon vertex 
function -- together with truncated perfect free quarks --
to heavy quarks is presently under investigation.
Preliminary results for the charmonium spectrum are given
in Ref. \cite{vert}.

The purpose of this paper is to test specifically such a
direct application in a very simple situation. Our model
is the 1d $\lambda \phi^{4}$ model, or anharmonic oscillator.

As a toy model, the anharmonic oscillator has a number of
virtues: we can achieve an excellent locality, such that
our $O(\lambda )$ perfect action does not need any truncation
of the couplings. Thus the perturbative improvement can
be tested separately, without admixture of truncation effects.
Moreover, our construction is based on continuum perturbation
theory, and there we do not encounter any divergent loop 
integrals, in contrast to field theory ($d>1$). Finally,
the reduction to quantum mechanics has the advantage that 
the couplings we identify do not get renormalized
in the full theory.

On the other hand, it is exceptionally difficult in our model
to demonstrate an improvement compared to the standard action,
because the latter is also very good in this case. For the harmonic
oscillator it is even perfect itself, for small interaction
-- the regime of interest here -- it is still very good,
and even for moderate interactions it performs amazingly well.
As a further problem we note that the performance of
continuum perturbation theory, which our improvement is based on, 
is rather poor in this model.

The advantages and disadvantages listed above are specific for
the one dimensional case.

\section{The model in the continuum}

The observables we are going to consider can be evaluated
directly in the continuum to a fantastic accuracy.
Our interest is of course not in their values, but solely in the
comparison of lattice artifacts in
different discretizations. We want to test the 
success of a specific improvement program for the lattice
action.

Nevertheless we have to start by recalling some properties of the
conti-nuum system. To fix the (field theoretic) notation, we denote the
Euclidean action as
\begin{equation} \label{contact}
s[\vp ] = \int dt \ \Big[ \frac{1}{2} \dot \vp (t)^{2}
+ \frac{m^{2}}{2} \vp (t)^{2} + \lambda \vp (t)^{4} \Big] .
\end{equation}
Throughout this paper we assume $m,\ \lambda \geq 0$, hence we
only study the ``symmetric phase'' 
(as opposed to the double well).
We consider the energy eigenvalues $E_{n}$, 
more precisely we are going to measure directly the energy gaps 
$\Delta E_{n} \doteq E_{n}-E_{0}$. 
An additive constant all over the
spectrum is out of control, and not much of interest either.
The simplest scaling quantity 
is
\begin{equation} \label{scal}
\frac{\Delta E_{2}}{\Delta E_{1}} \equiv \Delta E_{2} \cdot \xi \ ,
\end{equation}
where $\xi$ is the correlation length. Moreover,
there is the simple relation
\begin{equation} \label{rescale}
\frac{\Delta E_{n}(\mu^{2}m^{2}, \mu^{3} \lambda )}
{\Delta E_{n}(m^{2},\lambda )} = \mu \ , \quad {\rm any~}\mu >0 \ .
\end{equation}
This just follows from rescaling $t \to t /\mu$ and momentum
$k \to \mu k$ in the Hamiltonian ({\em not} rescaling 
all dimensional quantities).
\footnote{This argument was made rigorous first by K. Symanzik
(unpublished). The point is that the rescaling can be implemented 
unitarily \cite{Ban}.}

The quantity (\ref{scal}) is a {\em scaling quantity} in the strict sense,
i.e. a dimensionless ratio of physical observables.
On the other hand, quantities like (\ref{rescale}) may
also involve unphysical normalization factors. In a field theoretic
language they correspond to  the {\em asymptotic scaling}. Actually,
improved actions are designed for an improvement of scaling,
but the influence on asymptotic scaling is of interest too.
It has been observed before for the Gross Neveu model \cite{GN} and for
pure SU(3) gauge theory \cite{TdG} that ``accidentally'' the latter
is also improved for (quasi-)perfect actions. 

For $m>0$, the strength of the interaction depends on the 
dimensionless parameter
\begin{equation}
\tilde \lambda \doteq \frac{\lambda}{m^{3}} \ ,
\end{equation}
which is obvious from eq. (\ref{rescale}).
The energy eigenvalues can be expanded in $\tilde \lambda$,
but these expansions diverge at large orders
(the coefficients oscillate and their absolute values grow 
faster than any polynomial).
\footnote{Note that the point $\tilde \lambda =0$
is non analytic. For a discussion of large orders, see
e.g. Ref. \cite{BeWu}.}
However, a truncated series is still useful at
$\tila <<1$ (this situation is familiar from QED).
\footnote{We expect the same behavior also for the the couplings in
the perturbatively perfect action.}
The coefficients of these expansions have been derived many
times in the literature, for instance in Ref. \cite{expa},
\begin{eqnarray}
\frac{\Delta E_{1}}{m}(\tila ) & \simeq &
1 + 3 \tilde \lambda - 18 \tila^{2} + \frac{1791}{8} \tila^{3}
- 3825 \tila^{4} , \nonumber \\
\frac{\Delta E_{2}}{m}(\tila ) & \simeq &
2 + 9 \tilde \lambda - \frac{297}{4} \tila^{2} + 
\frac{9873}{8} \tila^{3} - \frac{1772685}{64} \tila^{4} ,
\nonumber \\ \label{expand}
\frac{\Delta E_{2}}{\Delta E_{1}} (\tila ) & \simeq &
2 + 3 \tilde \lambda - \frac{189}{4} \tila^{2} + 
\frac{7857}{8} \tila^{3} - \frac{1569069}{64} \tila^{4} \ . 
\end{eqnarray}
Tables of explicit values at finite $\tila$ are given for example
in Refs. \cite{Ban,Bis,BBCK}.
However, in particular for small $\tila$ they can easily be reproduced
from an eigenvalue problem, as described for instance in
Ref. \cite{Bis}.

\begin{figure}[hbt]
\vspace{-5mm}
\begin{center}
\epsfig{file=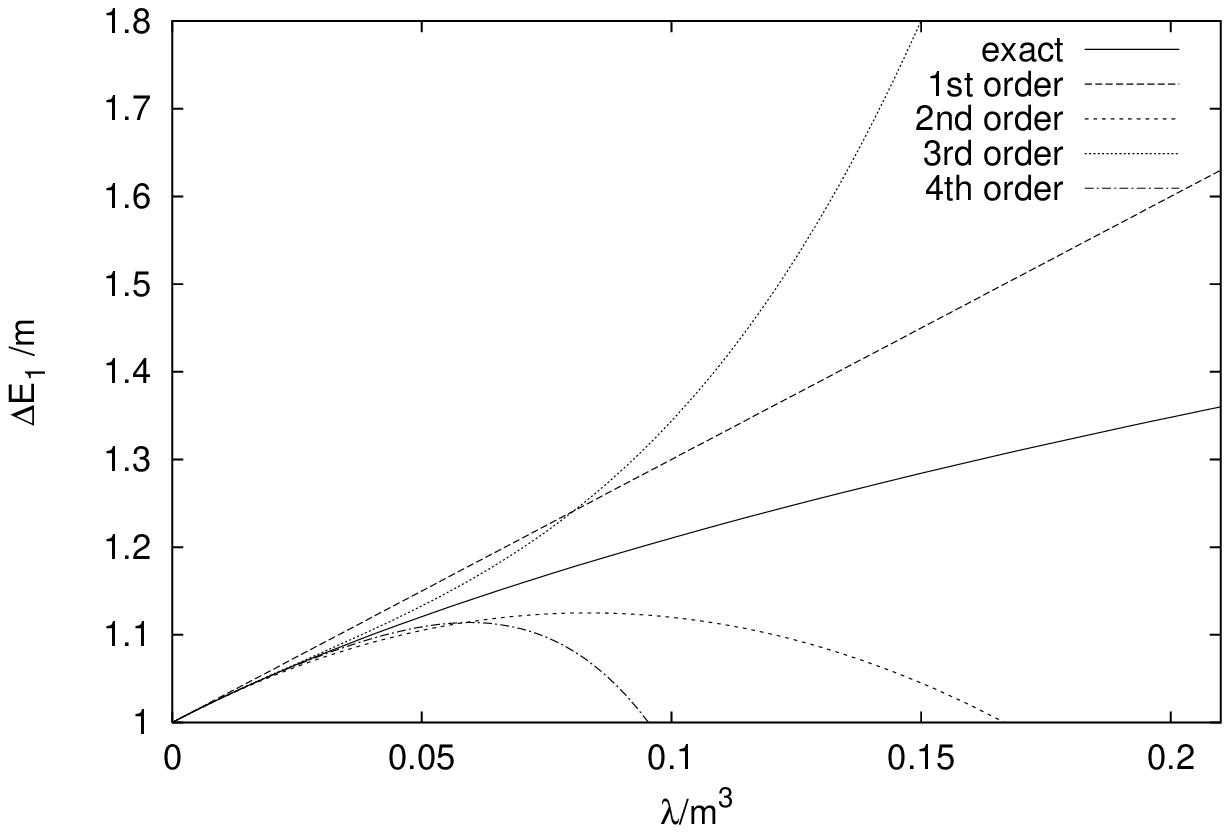,width=8.5cm,angle=0}
\epsfig{file=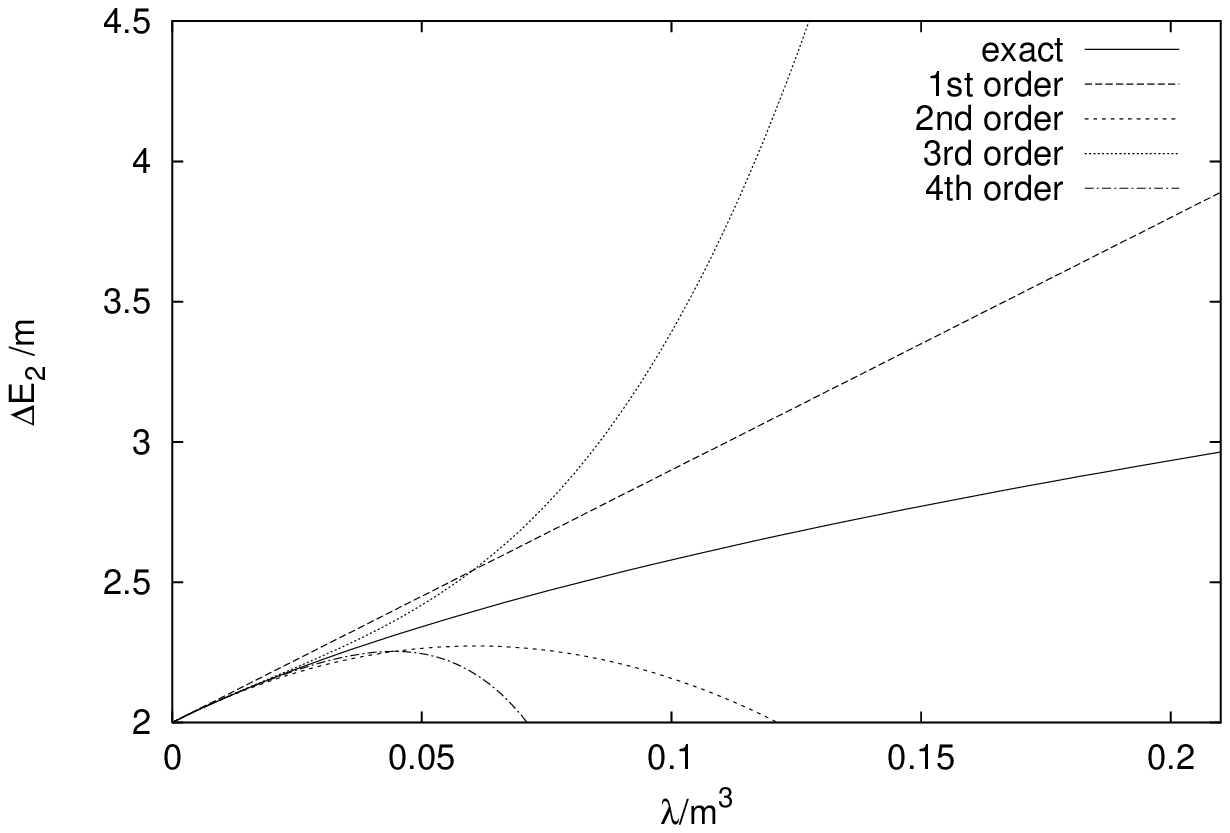,width=8.5cm,angle=0}
\end{center}
\caption{\it The ratios $\Delta E_1 /m$ (on top) and
$\Delta E_2 /m$ (below) as functions of
$\tila$. The exact result is compared to the perturbation
series truncated at various orders.}
\label{figE1}
\end{figure}

\begin{figure}[hbt]
\begin{center}
\epsfig{file=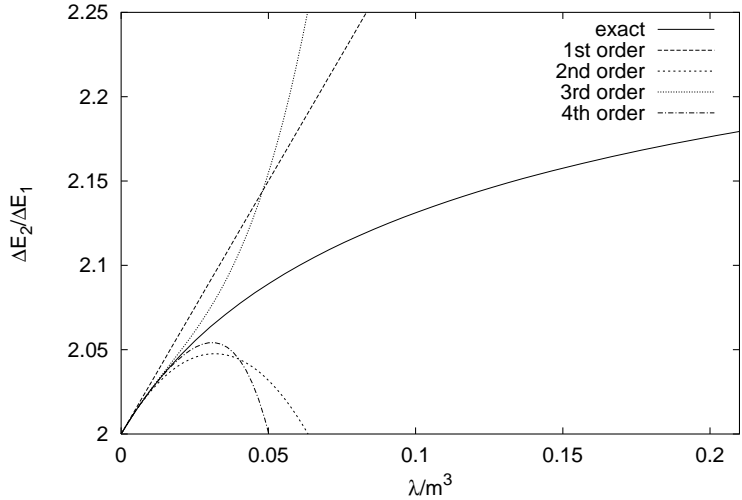, width=10cm,angle=0}
\end{center}
\caption{\it The ratio $\Delta E_2 / \Delta E_1$ as a function of
$\tila$. The exact result is compared to the perturbation
series truncated at various orders.}
\label{figE21}
\end{figure}

Since our construction in the following sections is also perturbative in
$\tila$, it is important to know how the perturbation series for
the above quantities behave.
Figs. \ref{figE1}
and \ref{figE21}
compare the exact function to the truncated
expansion in first, second, third and fourth order.




We see that the applicability of truncated expansions is restricted
to very small values of $\tila$. The exact range depends on the
quantity considered; it agrees with the relative magnitude of the
coefficients in the expansions (\ref{expand}).
This range gradually expands as we proceed to higher orders.

\section{The $O(\lambda )$ perfect action in momentum space}

If a system is given by some lattice action, then its
physical properties remain unaltered under a block
variable renormalization group transformation (RGT) \cite{WilKog}.
For suitable RGT parameters and infinite correlation length,
an infinite number of iterations may lead to a finite fixed 
point action (FPA). A FPA is an example for a perfect action,
since it is invariant under an RGT, and hence insensitive
to the lattice spacing. Perfect actions also exist at any
finite correlation length \cite{WilKog}. They reveal
exact continuum scaling at any lattice spacing.
For free or 
perturbatively interacting fields, they can be computed
analytically in momentum space. This calculation
simplifies if we send the blocking factor to infinity,
which amounts to a technique that we call ``blocking
from the continuum''. Recently this method has been applied
to the Schwinger model \cite{Schwing} and to QCD \cite{QuaGlu}.
Here we want to apply it to construct a lattice action for
the anharmonic oscillator, which is perfect to $O(\tila )$.
Our blocking uses the standard, piece-wise constant
weight distribution for the original variables, and a Gaussian
transformation term. For an alternative ansatz, closer to the 
spirit of a decimation RGT, see Ref. \cite{NKF}.

Our perfect lattice action $S[\phi ]$ is determined by the
functional integral
\begin{eqnarray}
e^{-S[\phi ]} &=& \int D\vp 
\exp \{ -s[\varphi ]-R[\phi ,\varphi ] \} \ , \nonumber \\
R[\phi ,\varphi ] &=& \frac{1}{2 \alpha} \sum_{x \in \Z} \Big(
\phi_{x} - \int_{x-1/2}^{x+1/2} dt \ \varphi (t) \Big)^{2} \ .
\end{eqnarray}
Here $\phi$ is the lattice field, $x$ are the lattice
sites and the continuum action $s[\varphi ]$ is given
in eq. (\ref{contact}). The RGT parameter $\alpha > 0$
is arbitrary; for any value of $\alpha$ the RGT keeps 
the partition function invariant,
\begin{equation}
Z = \int D \vp \ e^{-s[\vp ]} \propto \int D \phi \ e^{-S[\phi ]},
\quad (D\phi \doteq \prod_{x\in \Z} \int d\phi_{x}),
\end{equation}
and with it all expectation values.
The limit $\alpha \to 0$ corresponds to the well known
``$\delta$ function RGT''.

In momentum space, this expression can be written as
\begin{eqnarray}
e^{-S[\phi ]} &=& \int D\vp D \sigma \nonumber
\exp \Big\{ -\frac{1}{2\pi} \int_{-\pi}^{\pi} dk \ \times \\
&& \Big[ \frac{1}{2} \sum_{l\in \Z} \vp (-k-2\pi l)
[(k+2\pi l)^{2}+m^{2}] \vp (k+2\pi l) \nonumber \\
&& + i \sigma (-k)[\phi (k) - \sum_{l\in \Z}\vp (k+2\pi l)
\Pi (k+2\pi l)] \nonumber \\
&& + \frac{1}{2} \alpha \ \sigma (-k) \sigma (k) \Big] \Big\}
\times \nonumber \\
&& \Big\{ 1 - \frac{\lambda}{(2\pi )^{3}} \int d^{4}p \
\vp (p_{1}) \vp (p_{2}) \vp (p_{3}) \vp (p_{4}) \
\delta (\sum_{i=1}^{4}p_{i}) \nonumber \\
&& + O(\lambda^{2}) \Big\} , \nonumber \\
&& \Pi (k) \doteq \frac{\hat k}{k} \ , \quad
\hat k \doteq 2 \sin \frac{k}{2} \ ,
\end{eqnarray}
where we have introduced an auxiliary lattice field $\sigma$.

We denote  the free continuum propagator as
\begin{equation}
\Delta (k) = \frac{1}{k^{2}+m^{2}} ,
\end{equation}
and
\begin{equation}
G(k) = \sum_{l\in \Z} \Delta (k+2\pi l) \Pi(k+2\pi l)^{2}+\alpha
\end{equation}
is the perfect free lattice propagator, as we will
see. At $m=0$ this is the fixed point propagator, which has been
calculated for free scalar theories first by
Bell and Wilson, and it characterizes the FPA
for the O($N$) model in the large $N$ limit as well \cite{HS}.

We now choose the special value $\alpha =
(\sinh m - m)/m^{3}$, which renders the free lattice action
``ultralocal'' \cite{Anton}, i.e. it only couples nearest neighbor 
lattice variables,
\footnote{This choice for the RGT parameter
$\alpha$ also provides optimal locality in $d=4$ \cite{WB}.}
\begin{equation} \label{nn}
G(k) = \frac{\sinh m \cdot \hat m^{2}}{m^{3}}
\frac{1}{\hat k^{2} + \hat m^{2}} \ , \quad
\hat m \doteq 2 \sinh \frac{m}{2} .
\end{equation}

Our first step is the substitution
\begin{equation}
\tilde \vp (k+2\pi l) \doteq \vp (k+2\pi l) - i \sigma (k)
\Delta (k + 2\pi l) \Pi (k+2\pi l),
\end{equation}
which allows us to integrate out the continuum variable $\tilde \vp$.
We omit the constant factor in the Gaussian integral
\footnote{This is an example of an uncontrolled additive constant
in $S[\phi ]$, which motivates the consideration of the energy
{\em gaps}, rather than the single eigenvalues.
The same holds for the subsequent integration over $\tilde \sigma$,
see below.}
and obtain
\begin{eqnarray}
e^{-S[\phi ]} &=& \int D \sigma \exp \Big\{ - \frac{1}{2\pi }
\int_{-\pi}^{\pi} dk \ [i\sigma (-k) \phi (k) +
\frac{1}{2} \sigma (-k) G(k) \sigma (k)] \Big\} 
\times \nonumber \\
&& \Big\{ 1 - 3 \lambda \Big( \frac{1}{2m} \Big)^{2} \nonumber \\
&& + 6 \lambda \frac{1}{2m}  \Big[ \prod_{i=1}^{2}
\sum_{n_{i}\in \Z} \frac{1}{2\pi} \int_{-\pi}^{\pi}
dp_{i} \ \sigma (p_{i}) \Delta (p_{i}+ 2\pi n_{i}) 
\Pi (p_{i}+2\pi n_{i}) \Big] \nonumber \\
&& \qquad \times 2 \pi \delta (p_{1}+p_{2}) \delta_{n_{1},-n_{2}}
\nonumber \\
&& - \lambda \Big[ \prod_{i=1}^{4} \sum_{n_{i}} \frac{1}{2\pi}
\int_{-\pi}^{\pi} dp_{i} \ \sigma (p_{i}) \Delta (p_{i}
+ 2\pi n_{i}) \Pi (p_{i}+2\pi n_{i}) \Big] \nonumber \\
&& \qquad \times 2 \pi \delta ( \sum_{i=1}^{4}[p_{i}+2\pi n_{i}])
+ O(\lambda^{2})\Big\} .
\end{eqnarray}
In a sense, this computation goes beyond the perfect QCD
vertex function of Ref. \cite{QuaGlu}, because it includes
-- for the first time in the construction of a perfect
action -- a loop calculation, i.e. a {\em quantum correction}.
The continuum loop integral reads
\begin{equation}
\Delta (x)\vert_{x=0} = \frac{1}{2\pi} \int dk \
\Delta (k) = \frac{1}{2m}  \qquad (m>0),
\end{equation}
which has been inserted above. In field theory we would
encounter divergences at this point, which could be regularized
by some standard technique in the continuum.
Here the expression is finite from the beginning, and we will
see that even the limit $m\to 0$ can safely been taken
at the end, when we identify the couplings in the $O(\tila )$
perfect lattice action.

After performing a second substitution,
\begin{equation}
\tilde \sigma (k) \doteq \sigma (k) + iG(k)^{-1} \phi (k),
\end{equation}
we can integrate $\tilde \sigma$, and we arrive at a lattice
action of the form
\begin{eqnarray}
S[\phi ] &=& \frac{1}{2\pi} \int_{-\pi}^{\pi} dk \ \frac{1}{2}
\phi (-k) G(k)^{-1} \phi (k) \nonumber \\
&& + \lambda \Big[ A + \frac{1}{2\pi} \int_{-\pi}^{\pi}
dk \ \phi (-k) B(k) \phi (k) \nonumber \\
&& + \frac{1}{(2\pi )^{3}} \int_{-\pi}^{\pi} d^{4}p \
C(p) \phi (p_{1}) \phi (p_{2}) \phi (p_{3}) \phi (p_{4})
\Big] + O(\lambda^{2}) . \label{form}
\end{eqnarray}
This confirms that $G$ is the perfect free lattice
propagator for the RGT chosen here. Since it differs
from the standard propagator only by a constant factor and
a transformation of the mass, we can also confirm the
statement that for the {\em harmonic} oscillator the
standard action is perfect already.
\footnote{
In fact, any 
lattice action for the harmonic oscillator is perfect \cite{HN}.}
This is very specific for the case $d=1$ considered here.

The functions $B(k)$ and $C(p)=C(p_{1},p_{2},p_{3},p_{4})$ 
represent additional 2-variable and 4-variable
couplings, while $A$ is a constant, which is not really
of interest to $O(\lambda )$.

The Wick contraction of two isolated $\tilde \sigma$
variables yields the lattice loop integral
\begin{eqnarray}
\gamma (m) &\doteq & \frac{1}{2\pi} \sum_{l \in \Z}
\int_{-\pi}^{\pi} G(k)^{-1} \Delta (k + 2\pi l)^{2}
\Pi (k+2\pi l)^{2} \nonumber \\ \nonumber
&=& \frac{1}{m\cdot \sinh m \cdot \hat m^{2}}
\Big[  2 \cosh m + e^{-m} (1+\sinh m)- \frac{3}{m} \sinh m \Big] \\
&=& \frac{1}{2m} - \frac{7}{30} + \frac{11}{630}m^{2} + O(m^{4}) ,
\end{eqnarray}
which obeys $\gamma (m) - \gamma (-m) = 1/m$.

If we just insert this everywhere, we obtain the following
$O(\lambda)$ terms,
\begin{eqnarray}
A_{0} &=& 3 \Big[ \frac{1}{2m} - \gamma (m) \Big]^{2} \nonumber \\
B_{0}(k) &=& 3 \Big[ \frac{1}{m} - 2\gamma (m)\Big]
G(k)^{-2} \sum_{l\in \Z} \Delta (k+2\pi l)^{2}
\Pi (k+2\pi l)^{2} \nonumber \\
&=& \Big[ \frac{1}{m} - 2 \gamma (m) \Big] \frac{3m^{2}}
{2 (\sinh m \cdot \hat m^{2})^{2}} \times \nonumber \\
&& \Big\{ \hat k^{4}(\frac{1}{2} \hat m^{2} + \tilde m )
+ \hat k^{2} \hat m^{2} \tilde m + 2 \hat m^{4} \Big\}
\nonumber \\
{\rm where} && \tilde m \doteq 3 \Big( 1 - \frac{\sinh m}{m}
\Big) \nonumber \\
C(p) &=& \Big[ \prod_{i=1}^{4}
G(p_{i})^{-1} \sum_{n_{i}\in \Z} \Delta (p_{i}+2\pi n_{i})
\Pi (p_{i}+2\pi n_{i})\Big] \times \nonumber \\
&& \delta ( \sum_{i=1}^{4} [p_{i} + 2\pi n_{i}]) \ . \label{naco}
\end{eqnarray}

Note, however, that in the contractions in the $\sigma^{4}$ term,
an expectation value
$\langle \tilde \sigma (p_{i}) \tilde \sigma (p_{j}) \rangle$
only enforces $p_{i}=-p_{j}$ (`pairing' of the lattice momenta), 
while the related summation integers $n_{i},\ n_{j}$ 
remain independent (no `pairing' of the continuum momenta).
Therefore, such contractions yield an additional contribution
$S(k), ~ (k \in ]-\pi ,\pi ]) $, which does not factorize,
\begin{eqnarray}
S(k) & \doteq & \frac{1}{2\pi} \int_{-\pi}^{\pi} dq \
G(q)^{-1} \Big[ \sum_{n_{1},n_{2},n_{3},n_{4}} \Delta (q+2\pi n_{1})
\Pi (q + 2\pi n_{2}) \times \nonumber \\
&& \Delta (-q+2\pi n_{2}) \Pi (-q + 2\pi n_{2})
\Delta (k+2\pi n_{3}) \Pi (k + 2\pi n_{3}) \nonumber \\
&& \Delta (-k+2\pi n_{4}) \Pi (-k + 2\pi n_{4}) \Big] \
\delta_{\sum_{i}n_{i},0} \ (1-\delta_{n_{1},-n_{2}}) . \label{sdef}
\end{eqnarray}
(The case $n_{1}+n_{2}=0=n_{3}+n_{4}$ is excluded here, because it has
been included before in eq. (\ref{naco}).)
This modifies the constant and the bilinear term to
\begin{eqnarray} \nonumber
A &=& A_{0}+ \frac{3}{2\pi} \int_{-\pi}^{\pi} dk \ G(k)^{-1}S(k), \\
B(k) &=& B_{0}(k) - 6 G(k)^{-2}S(k),
\end{eqnarray}
while the vertex function $C(p)$, given in eq. (\ref{naco}),
is not affected.
It does not pick up any loop contributions, and its structure
can easily be understood in the language of ``building blocks''
as introduced for the quark-gluon vertex \cite{StL}.

The lattice spacing artifacts in the standard action can be
classified in magnitudes as $\lambda a^{2},\ \lambda a^{4},
\ \lambda a^{6} ,\dots ,\ \lambda^{2} a^{2}, \lambda^{2} a^{4},
\dots , \lambda^{3} a^{2} \dots $. In the action we have constructed
now, all artifacts $\propto \lambda $ are erased. The remaining 
artifacts can still be $\propto a^{2}$, but they are multiplied
at least by $\lambda^{2}$. This is analogous to QCD with the perfect
vertex function, where the gauge coupling $g$ plays the r\^{o}le
of $\lambda$. There the artifacts of the Wilson action start even
in $O(ga)$, and the perturbative perfection pushes the leading
artifact to $O(g^{2}a)$.

\section{The perfect couplings in coordinate space}

The interaction terms involving $B$ and $C$, which we derived
in momentum space, turn into convolutions in coordinate space.
Hence $B(r)$ describes additional 2-variable couplings
(``2-spin couplings'' in a solid state language)
over a distance $r\in \Z$, and $C(r_{1},r_{2},r_{3},0)
\doteq C(\vec r )$ introduces 4-spin couplings,
\begin{displaymath}
\lambda \sum_{x,r \in \Z} B(r) \phi_{x} \phi_{x+r} +
\lambda \sum_{x \in \Z,\vec r \in \Z^{3}} 
C(\vec r) \phi_{x+r_{1}} \phi_{x+r_{2}} \phi_{x+r_{3}} \phi_{x}.
\end{displaymath}
We exploit lattice
translational invariance, in the latter case by setting $r_{4}$
to the arbitrary value 0. $C(r_{1},r_{2}r_{3},0)$ is invariant
under permutation of its components, and $C(\vec r )= C(-\vec r)$
(but there is {\em no} invariance under sign flip of just one or two
components of $\vec r$).

\subsection{The 2-spin couplings}

The function $B(k)$ is even, hence
\begin{equation}
B(r) \doteq B_{0}(r)+B_{1}(r) =
\frac{1}{2\pi} \int_{-\pi}^{\pi} dk \ [B_{0}(k)
- 6 G(k)^{-2}S(k)] \cos (kr) .
\end{equation}
The first term, $B_{0}(r)$, can be computed analytically.
It turns out to be the dominating contribution to $B(r)$,
\begin{eqnarray}
B_{0}(r) &=& 3 \ \Big[ \frac{1}{2m}-\gamma (m) \Big] \
\frac{m^{2}}{(\sinh m \cdot \hat m^{2})^{2}} \times \nonumber \\
&& \Big\{ \beta_{0} \delta_{r,0} + \beta_{1} [\delta _{r,1}+
\delta_{r,-1}] + \beta_{2} [\delta_{r,2}+\delta_{r,-2}] \Big\} ,
\nonumber \\
\beta_{0} &=& 2 \ (3+\hat m^{2})\tilde m + 3\hat m^{2} + 2 \hat m^{4}
, \nonumber \\
\beta_{1} &=& -(4+\hat m^{2})\tilde m - 2 \hat m^{2} ,
\nonumber \\
\beta_{2} &=& \tilde m + \frac{1}{2} \hat m^{2}.
\end{eqnarray}
(The quantities $\hat m$ and $\tilde m$ are defined
in eqs. (\ref{nn}) and (\ref{naco}).)

Expanding in small $m$, we recognize the finiteness
of this expression in the limit $m\to 0$,
\footnote{The finiteness at $m=0$ (``quartic oscillator'')
is a very sensitive consistency test. We did not
study that case -- where $\tila = \infty$ and
$E_{n} = c_{n} \lambda^{1/3}$ -- extensively,
since our improved action is designed
for small $\tila$. However, we observed that $E_{0} \propto 
\lambda^{1/3}$ can be fitted better for the perturbatively perfect
action than for the standard action.} 
\begin{eqnarray}
B_{0}(r)&=& b_{0}^{(0)} \delta_{r,0} + b_{1}^{(0)}
[ \delta_{r,1}+\delta_{r,-1} ] + b_{2}^{(0)}
[ \delta_{r,2}+\delta_{r,-2} ] , \nonumber \\
b_{0}^{(0)} &=& \frac{77}{100} - \frac{419}{1400} m^{2}
+O(m^{4}) , \nonumber \\
b_{1}^{(0)} &=& \frac{91}{300} - \frac{1637}{12600} m^{2}
+O(m^{4}) , \nonumber \\
b_{2}^{(0)} &=& \frac{7}{600} - \frac{31}{5040} m^{2}
+O(m^{4}).
\end{eqnarray}

The additional term, $B_{1}(r)$, has to be evaluated numerically.
It is significantly suppressed, essentially because
the case $n_{i}=0, \ i=1 \dots 4$ is excluded from the summation
in eq. (\ref{sdef}). It typically affects the bilinear couplings
only in third digit.

\begin{figure}[hbt]
\begin{center}
\epsfig{file=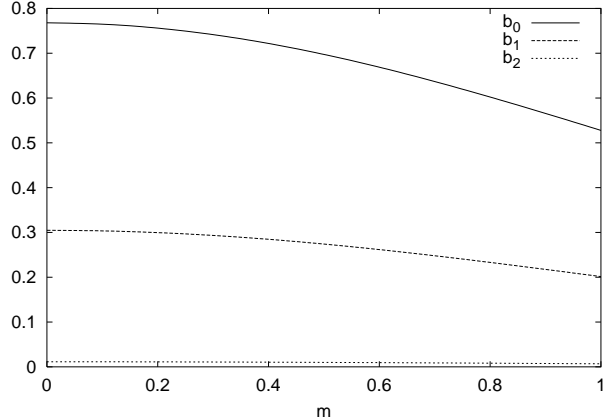,width=8.5cm,angle=0}
\end{center}
\caption{\it The perfect bilinear couplings $\propto \tila $,
as functions of the mass.}
\label{figc2}
\end{figure}

It turns out that also in $B_{1}(r)$ the couplings are restricted
to distances $\leq 2$. Therefore, the entire bilinear term 
$B(r)$ is given by $b_{i}=b_{i}^{(0)}+b_{i}^{(1)}$, $i=0,1,2$.
These couplings are shown as functions of the mass in
Fig. \ref{figc2}, and
some precise values are given Table \ref{tabcop}. 
For completeness we also include the constant $A$.


\begin{table}
\begin{center}
\begin{tabular}{|c|c|c|c|}
\hline
& $m=0$ & $m=0.2$ & $m=0.3$ \\
\hline
\hline
$A$       & 0.1634448589 & 0.1627118789 & 0.1615030089 \\
\hline
$b_{0}$   & 0.7648784176 & 0.7560248574 & 0.7415398190 \\
$b_{1}$   & 0.3034599619 & 0.2995928547 & 0.2932745430 \\
$b_{2}$   & 0.0112505782 & 0.0110757933 & 0.0107911914 \\
\hline
$C_{0}$   & 0.2242402815 & 0.2202292735 & 0.2137198561 \\
$C_{1}$   & 0.0579948140 & 0.0568473063 & 0.0549896682 \\
$C_{2}$   & 0.0002458836 & 0.0002396038 & 0.0002295148 \\
$C_{11}$  & 0.0386197986 & 0.0378246484 & 0.0365388182 \\
$C_{12}$  & 0.0008364070 & 0.0008162208 & 0.0007837264 \\
$C_{22}$  & 0.0000983088 & 0.0000956674 & 0.0000914309 \\
$C_{112}$ & 0.0042341197 & 0.0041403776 & 0.0039890713 \\ 
\hline
\end{tabular}

\vspace{8mm}

\begin{tabular}{|c|c|c|c|}
\hline
& $m=0.4$ & $m=0.5$ & $m=1$ \\
\hline
\hline
$A$       & 0.1598369028 & 0.1577387270 & 0.14198250413 \\
\hline
$b_{0}$   & 0.7218143338 & 0.6973665904 & 0.52780256487 \\
$b_{1}$   & 0.2846878032 & 0.2740741694 & 0.20145092650 \\
$b_{2}$   & 0.0104063833 & 0.0099339708 & 0.00680856517 \\
\hline
$C_{0}$   & 0.2049621122 & 0.1942810514 & 0.12580113799 \\
$C_{1}$   & 0.0524997264 & 0.0494780115 & 0.03056403944 \\
$C_{2}$   & 0.0002161447 & 0.0002001630 & 0.00010708640 \\
$C_{11}$  & 0.0348181119 & 0.0327344171 & 0.01982579170 \\
$C_{12}$  & 0.0007405382 & 0.0006887131 & 0.00038135880 \\
$C_{22}$  & 0.0000858306 & 0.0000791587 & 0.00004091627 \\
$C_{112}$ & 0.0037871600 & 0.0035435707 & 0.00206185706 \\ 
\hline
\end{tabular}
\end{center}
\caption{\it The $O(\lambda )$ perfect couplings for masses 
$0 \dots 1$.}
\label{tabcop}
\end{table}

In addition we have of course the bilinear couplings of the free
theory (resp. harmonic oscillator), given in eqs. (\ref{nn})
and (\ref{form}). In coordinate space the free action reads
\begin{equation}
S[\phi ]_{\lambda =0} = \frac{m^{3}}{2 \sinh m \cdot \hat m^{2}}
\sum_{x,y \in \Z} \phi_{x} \Big[ (2+\hat m^{2})\delta_{x,y}
- \delta_{x,y+1} - \delta_{x,y-1} \Big] \phi_{y} \ .
\end{equation}

\subsection{The 4-spin couplings}

We recall that we describe the 4-spin couplings by $C(\vec r )$,
$\vec r = (r_{1},r_{2},r_{3}) \in \Z^{3}$, $r_{4}=0$,
that we have permutation invariance among $r_{1},\dots ,r_{4}$
and invariance under $\vec r \to -\vec r$, hence
\begin{equation}
C(\vec r ) = \frac{1}{(2\pi )^{3}} \int_{-\pi}^{\pi} d^{4}p\ 
C(p) \ \cos (\vec p \cdot \vec r ) \ .
\end{equation}

Note that all the singularities at $p_{i}=0$ are
removable, both, for finite and for vanishing mass.

\begin{figure}[hbt]
\vspace{-4mm}
\begin{center}
\epsfig{file=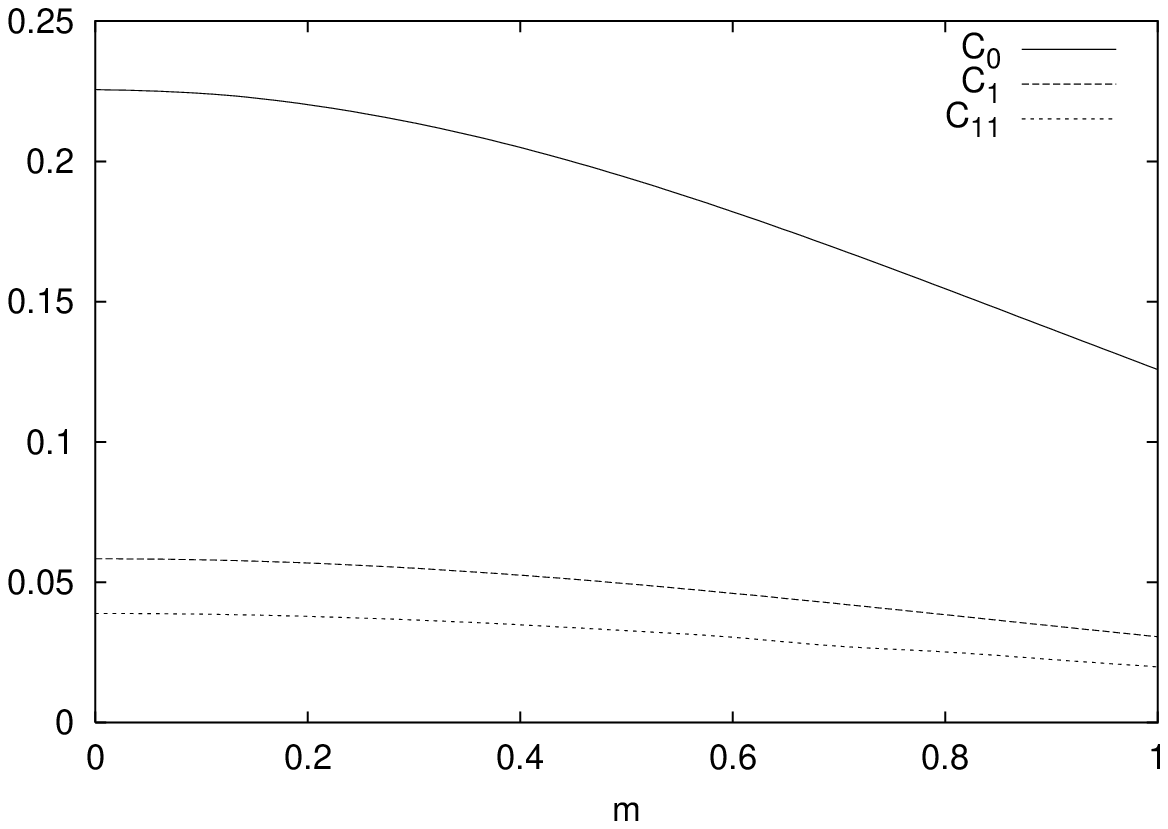,width=8.5cm,angle=0}
\epsfig{file=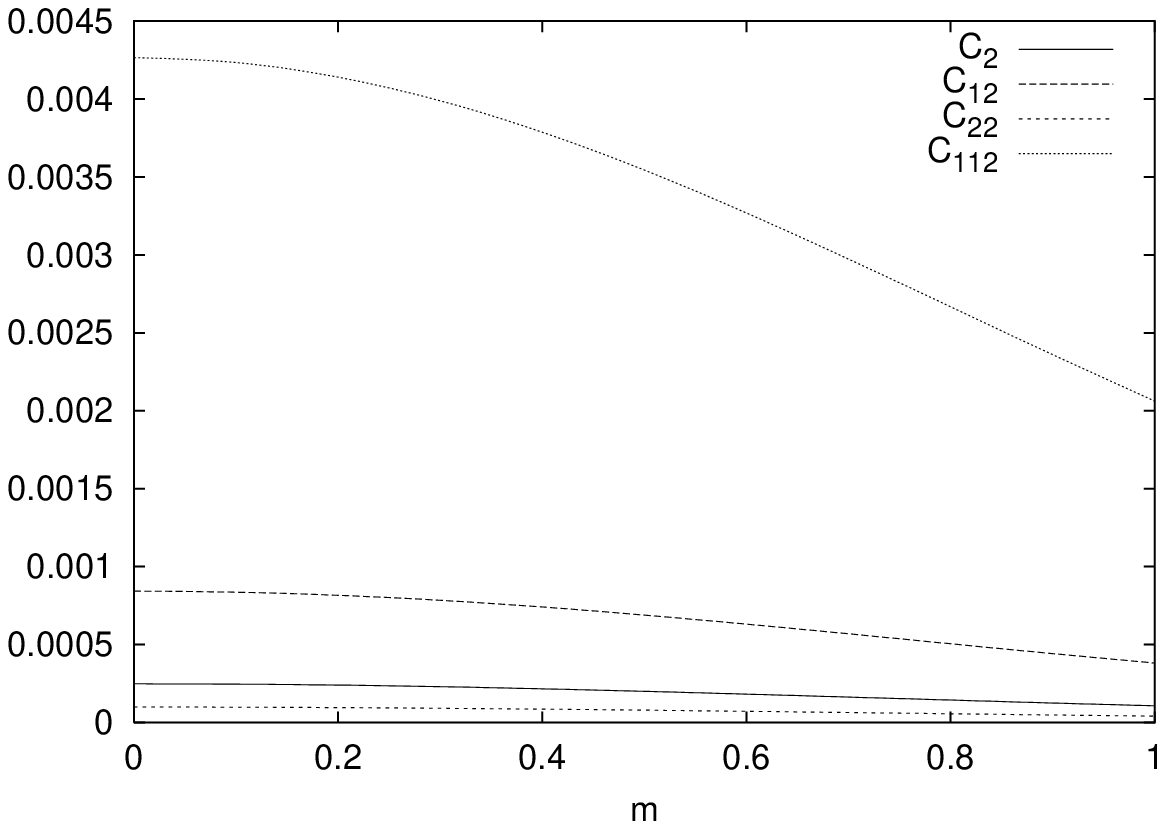,width=8.5cm,angle=0}
\end{center}
\vspace{-2mm}
\vspace{-3mm}
\caption{\it The perfect 4-spin couplings $\propto \tila $
as functions of $m$. 
Couplings involving distances $\leq 1$
are on top, couplings involving distance 2 below.}
\vspace{-3mm}
\label{figc4-1}
\end{figure}

It turns out that again the couplings never involve any two spins
separated by a distance larger than 2. A general argument
for that is given in the following subsection. This means
that there are just 7 independent 4-spin couplings. 
We denote them as
\begin{eqnarray*}
&& \hspace{-7mm} C_{0} \doteq C(\vec 0 ); \
C_{1} \doteq C(1,0,0) ; \
C_{2} \doteq C(2,0,0) \\
&& \hspace{-7mm} C_{11} \doteq C(1,1,0) ; \
C_{12} \doteq C(1,2,0) ; \
C_{22} \doteq C(2,2,0) ; \
C_{112} \doteq C(1,1,2) .
\end{eqnarray*}
They all represent equivalence
classes, which contain a total of 65 nontrivial couplings 
(keeping $r_{4}=0$ fixed). Some exact values are given 
in Table \ref{tabcop},
and their mass dependence is illustrated in 
Fig. \ref{figc4-1}.



\subsection{Locality}

Assume that we calculate the perfect action to $O(\lambda^{n})
\ (n \geq 1)$. This involves a number of perturbative correction
terms, which arise from the expectation values of
the continuum field $\vp$ to some power. The highest power
is $\langle \vp^{4n}\rangle $. There, Wick contractions
lead to several kinds of terms, which we classify by the
power of the inverse free propagator $G^{-1}$.
The maximal power is $G^{-2n}$.

For instance, in the bilinear term, which depends only on one
momentum, the maximal factor is $G(k)^{-2n}$.
Therefore the maximal power of $\hat k$ is
$\hat k^{4n} \propto (1-\cos k)^{2n}$, which can be decomposed
into terms $\propto \cos k, \dots ,\cos 2nk$.
In coordinate space this yields couplings $\propto
\delta_{x,1}+\delta_{x,-1}, \dots , 
[\delta_{x,2n}+\delta_{x,-2n}]$.

In the terms, which couple more than two lattice variables,
an analogous consideration leads to a `maximal'
factor $\prod_{i=1}^{2n} \cos p_{i}$, where the momenta $p_{i}$ 
may be all or partially different. The observation that variables
can not be coupled over distances $>2n$ still holds.

We have seen this restriction explicitly for $n=1$.
Since the couplings are confined to such a short range,
we can easily include all of them. Unlike field theory,
no truncation -- which does harm to the improved
properties of the action -- is needed. This allows us to study
the quality of a perturbatively improved action separately,
whereas in field theory one can only study a superposition
of the improvement and the truncation scheme.

We also note that -- within the limited set of couplings we
deal with -- locality becomes even better if the mass increases.
For instance, a larger mass suppresses the couplings over 
distance 2 even more, relative to the leading coupling constants
(see Table \ref{tabcop}).

The restriction of the couplings to a finite range is in
qualitative agreement with the effectively 1d quark-gluon
vertex function (quark fields constant in all but one direction)
\cite{QuaGlu}.
Moreover, also the increase of locality for rising mass
agrees with fermionic models and with scalar fields
in higher dimensions \cite{WB}.

\section{Numerical results for the energy gaps}

As a warming up exercise we computed the lattice partition function
from direct integration. If we do so at $\tila =0$ and some $\tila
>0$, we can extract an estimation for $E_{0}(\tila )-E_{0}(0)$.
In fact, this works much better for the perturbatively perfect action
than for the standard action; for instance at $m=\lambda =1$
the continuum  ground state energy $E_{0,cont}=0.8038$ is approximated
well using the improved action, 0.8021, whereas the standard action
yields 0.7111. However, this involves additive constants, which
may depend on $\tila$, so we focus on the energy {\em gaps} now.

\subsection{The simulation}

To compare the performance of the standard discretization and the
perturbatively perfect action, we simulated both actions
at correlation lengths
ranging from $\xi\approx2 \dots 5$, on a $L=30$ lattice with 
periodic boundary conditions,
using a standard Metropolis multi-hit algorithm. 
The first two energy gaps were extracted from the correlation functions 
$\langle 0|\phi (x) \phi (0)|0 \rangle $ and 
$\langle 0|\phi (x)^{2} \phi(0)^{2}|0 \rangle$, 
and the statistical errors 
were estimated by jackknife analysis.
The decay of the 
$\phi^{\ell}$ correlation function (in infinite volume) is given by
\footnote{At $L<\infty$ we actually obtain cosh functions, but we
can easily measure the decay in a region, where this difference is
negligible.}
\begin{equation}
\langle 0|\phi (x)^{\ell}\phi (0)^{\ell}|0 \rangle =
\sum_{n=0}^{\infty}
|\langle 0|\phi ^{\ell}|n \rangle |^{2}\exp(-\Delta E_{n}x),
\end{equation}
($\Delta E_{n} \doteq E_{n}-E_{0}$).
For $\ell =1,2$ it reduces to \footnote{Due to the mirror symmetry of the
potential, eigenfunctions for $E_n$ have parity $(-)^n$.} 
\begin{eqnarray} 
\langle 0|\phi (x) \phi (0)|0 \rangle & = & \sum_{n=0}^{\infty}
|\langle 0|\phi |2n+1 \rangle |^{2}\exp(-\Delta E_{2n+1}x ), \nonumber \\ 
\langle 0|\phi (x)^{2}\phi (0)^{2}|0 \rangle & = & \sum_{n=0}^{\infty}
| \langle 0|\phi^{2}|2n \rangle |^{2}\exp(-\Delta E_{2n}x) ,
\end{eqnarray} 
and for the harmonic oscillator we are left with 
\begin{eqnarray} 
\langle 0|\phi (x)\phi (0)|0\rangle & = & 
|\langle 0|\phi |1 \rangle |^{2}\exp(-\Delta E_{1}x), \nonumber \\
\langle 0|\phi (x)^{2}\phi (0)^{2}|0 \rangle &=&
|\langle 0|\phi^{2}|0\rangle |^{2} +
|\langle 0|\phi^{2}|2 \rangle |^{2} \exp(-\Delta E_{2}x). \label{decay}
\end{eqnarray}
In our simulations we study small interaction
parameters $\tila \leq 0.2$. In this regime,
varying the fitting ranges reveals that
the $\ell = 1,2$ correlation functions
do not pick up significant contributions from energy gaps higher
than the leading ones given in eq. (\ref{decay}). 

\subsection{Numerical results}

Simulations were done for masses and anharmonic 
couplings in the 
range $m = 0.2 \dots 0.5$ and $\tila = 0.001 \dots 0.2$.
We compared the first two energy gaps as ``asymptotic scaling 
quantities'', as well as $\Delta E_{2}/\Delta E_{1}$ as a 
scaling quantity.
For a direct evaluation, we divide the lattice results by
the corresponding continuum values.

\begin{figure}[hbt]
\def\fpsangle{0}
\epsfxsize=130mm
\fpsbox{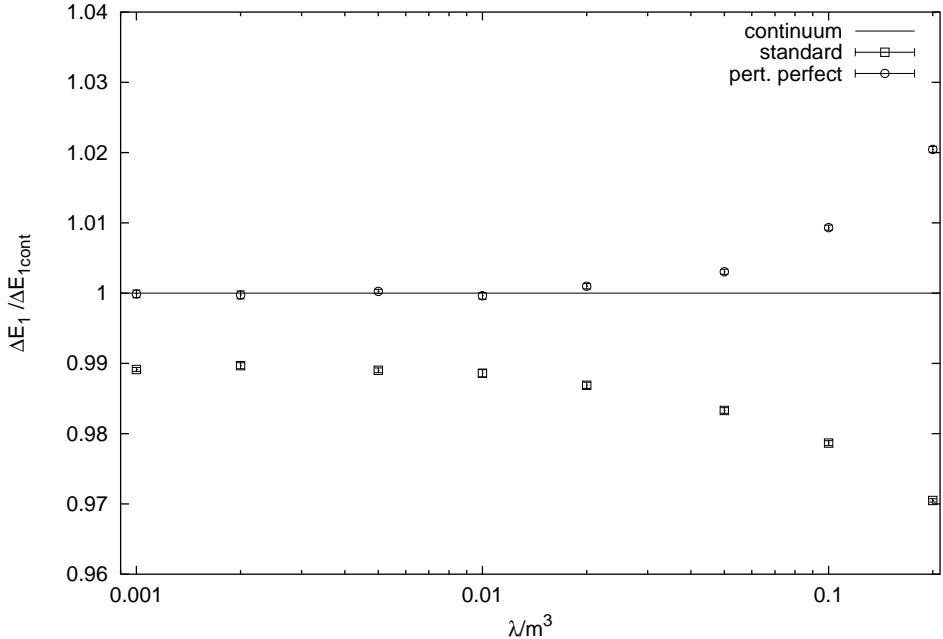}
\caption{\it The ratio $\Delta E_1 /\Delta E_{1,cont}$ at $m=0.5$,
as a function of $\tila$. }
\label{figp1}
\end{figure}

\begin{figure}[hbt]
\def\fpsangle{0}
\epsfxsize=130mm
\fpsbox{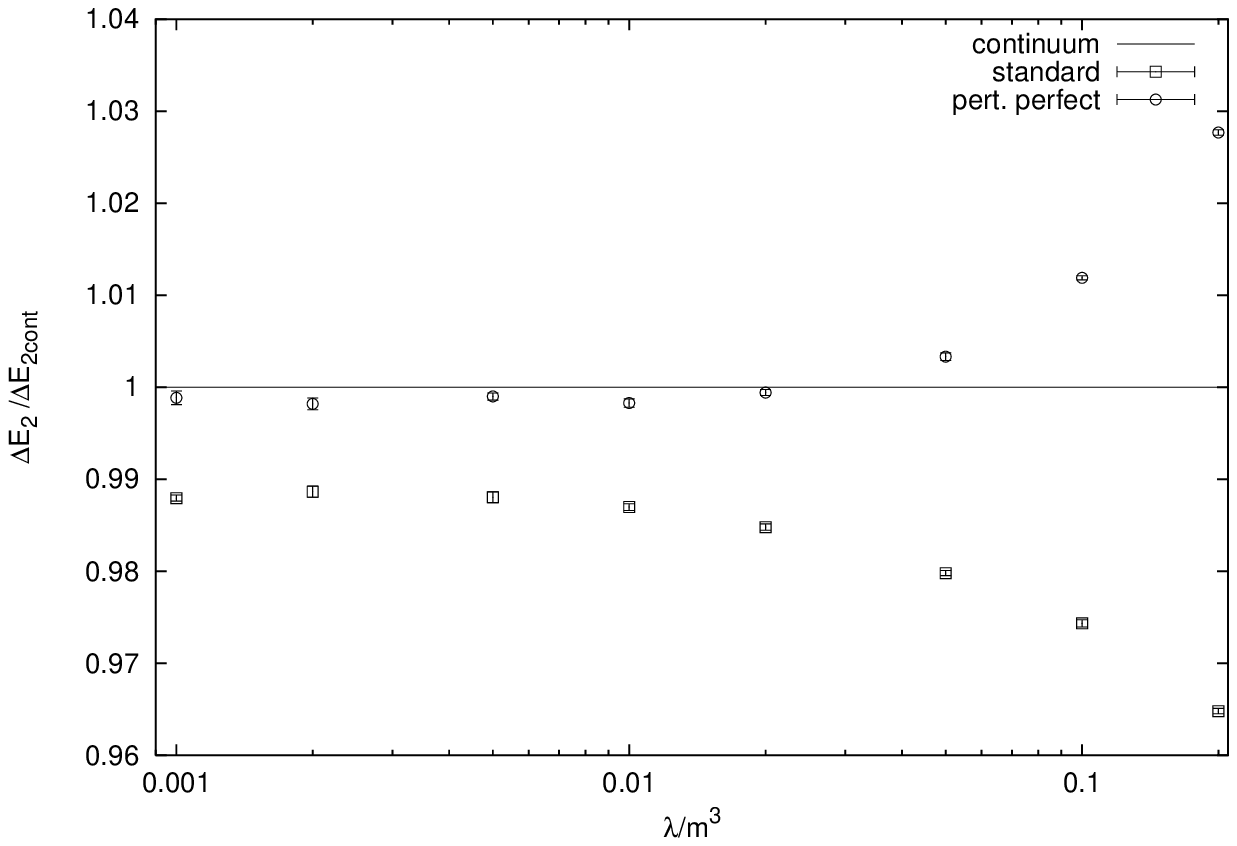}
\caption{\it The ratio $\Delta E_2 /\Delta E_{2,cont}$ at $m=0.5$,
as a function of $\tila$. }
\label{figp2}
\end{figure}

Figs. \ref{figp1} and \ref{figp2} show the gaps $\Delta E_{1} (\tila)$
and $\Delta E_{2}(\tila)$,  measured at $m=0.5$ for the 
standard and perturbatively perfect action 
(and normalized by their corresponding
continuum gaps $\Delta E_{n,cont}$).
These plots show that for anharmonic couplings
up to $\tila \approx 0.05$ the perturbatively perfect action
reproduces the 
continuum gaps much better than the standard action. Comparing
Figs. \ref{figp1}, \ref{figp2} with 
Fig. \ref{figE1},
one recognizes roughly the same reliability range for first order 
perturbation theory in the continuum.
This qualitative behavior of the improvement holds for a variety of
masses.

For the harmonic oscillator in infinite volume,
the perfect action reproduces the continuum gaps exactly, whereas
the standard action yields
\begin{equation}
\Delta E_{n,stan}(m) = n \cdot {\rm arccosh} (1 + m^{2}/2) .
\end{equation}
(The value $\Delta E_{1,stan}(m=0.5) \simeq 0.495$ agrees with Fig. 
\ref{figp1}).
The fact that even the standard action is perfect at $\tila =0$
is reflected by the exact values of the gap ratios.

\begin{figure}[hbt]
\def\fpsangle{0}
\epsfxsize=130mm
\fpsbox{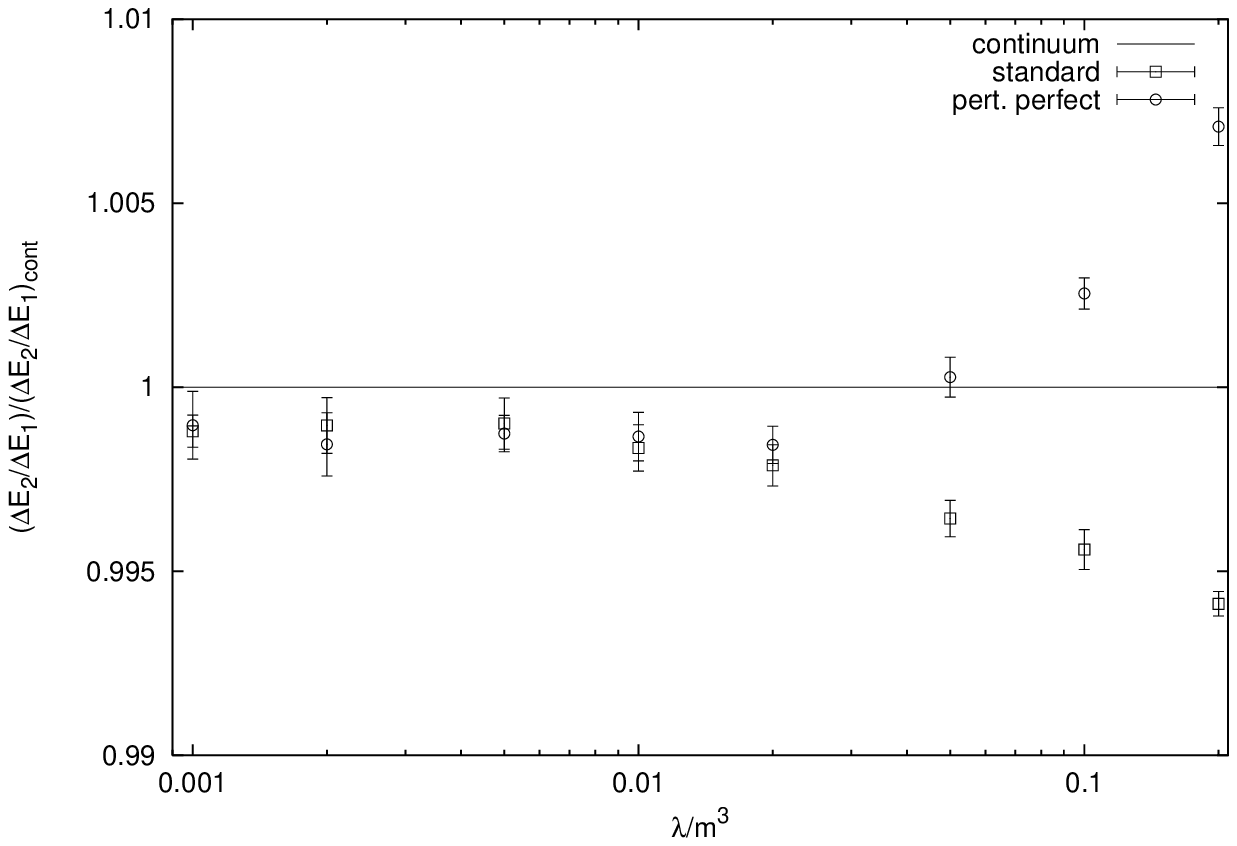}
\caption{\it The ratio $\frac{\Delta E_{2}}{\Delta E_{1}}/
\frac{\Delta E_{2}}{\Delta E_{1}}_{cont}$ at $m=0.5$,
as a function of $\tila$.}
\label{figp4}
\end{figure}

The scaling ratio $\frac{\Delta E_{2}}{\Delta E_{1}}(\tila)$ -- 
again measured at $m=0.5$ and normalized by
the continuum value --
is shown in Fig. \ref{figp4}.
Unfortunately there is hardly any conclusive difference between the 
two types of action for that quantity. 

\begin{figure}[hbt]
\def\fpsangle{0}
\epsfxsize=130mm
\fpsbox{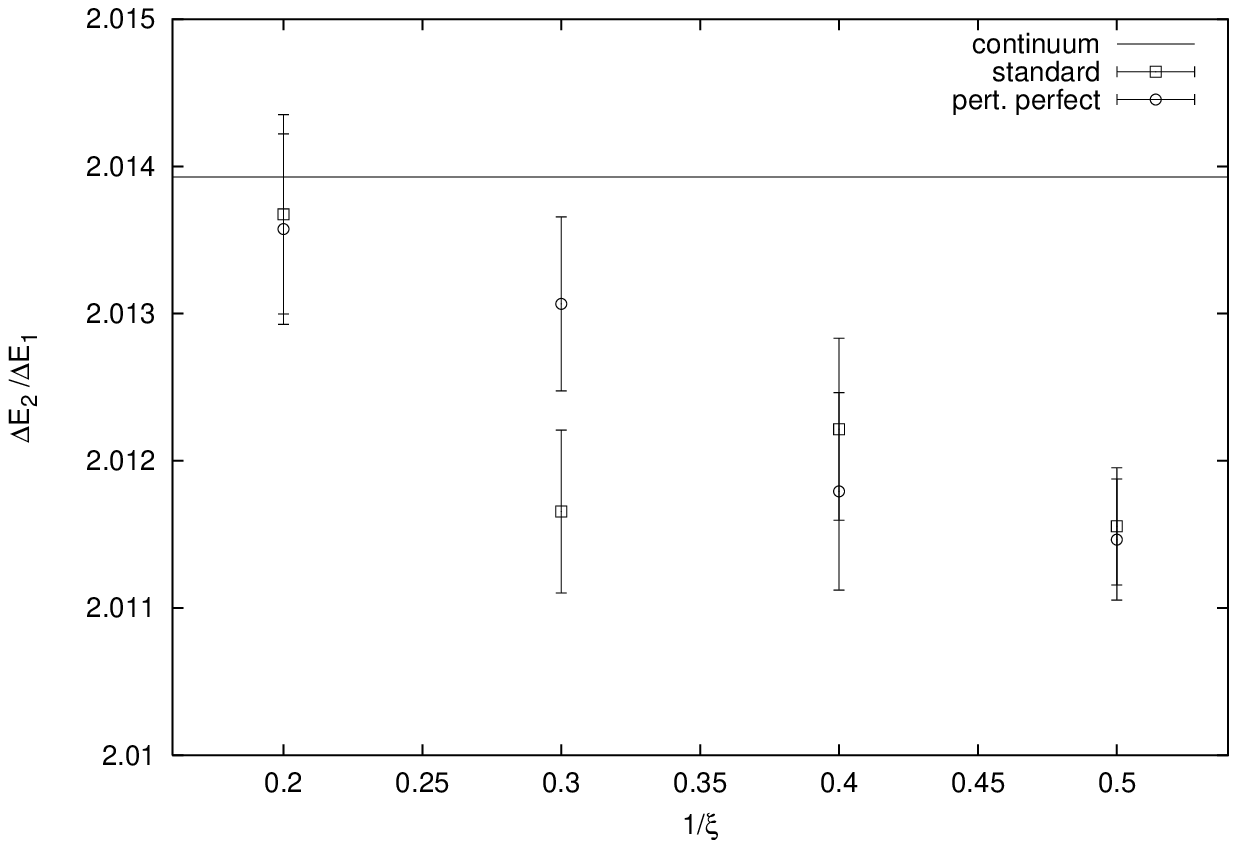}
\caption{\it The ratio $\Delta E_2 /\Delta E_1$ at fixed $\tila=0.005$.}
\label{figp5}
\end{figure}

Nevertheless, in Fig. 
\ref{figp5}, which shows the gap ratios versus $1/\xi$ 
at fixed $\tila=0.005$, a better performance 
of the perturbatively perfect action is visible at intermediate
correlation length. This is a scaling plot,
based on $10^{9}$ Monte Carlo sweeps.

As we know from Fig. \ref{figE21}, the applicability of first order
perturbation theory for $\Delta E_{2}/\Delta E_{1}$ is restricted
to really tiny values of $\tila$. The hope that the perturbatively
perfect action could still perform well beyond that range can not
be confirmed.
\footnote{We also looked at stronger interactions, such as
$\tila = 0.5$, for masses $m=0.3$ and $m=0.5$.
In that regime, the standard action scales even
better than the perturbatively perfect action.}
In that range itself, the improvement is extremely
difficult to demonstrate, because the standard action -- being perfect
at $\tila = 0$ -- is also excellent there. The available accuracy
is mainly limited by the fitting precision of the exponential
decays.

\section{Conclusions and outlook}

We have constructed a lattice action for the anharmonic
oscillator, which is perfect to first order in perturbation
theory. This is the first manifestly one loop perfect lattice
action
\footnote{For some time it was claimed that for asymptotically
free theories, FPAs are not only classically perfect, but 
automatically also one loop quantum perfect.
However, this claim has recently been disproved \cite{HN}.}.
Comparing this action to the standard lattice formulation,
we observe a clear improvement for the energy gaps
$\Delta E_{1}$, $\Delta E_{2}$ up to $\tila \doteq 
\lambda /m^{3} \sim 0.2$, over a variety of correlation lengths.
As a consequence, (pseudo-)scaling laws of the type of eq. 
(\ref{rescale}) are improved at small interactions.
In field theory, this corresponds to an improved asymptotic
scaling. Unfortunately, for the scaling quantity $\Delta E_{2}/
\Delta E_{1}$ an improvement could only be demonstrated
laboriously.
It seems to be restricted to very small values of $\tila$, in 
agreement with the performance of continuum perturbation
theory. There the linear approximation is useful
only for $\tila \leq O(10^{-2})$.
One could have hoped that the improved action
is successful also beyond this regime, but it turned out
that this is not the case. This can be viewed as a negative sign for
the direct application of perturbatively perfect actions,
but the outcome might of course depend on the model.
As a further test one could simulate the perturbatively classically
perfect action for the 2d O(3) model, which has been presented
-- but not tested -- in Ref. \cite{Has} (Table 2).

Moreover, in the regime of tiny $\tila$,
the standard action is also exceptionally successful in this toy model. 
Therefore, an accuracy of 4 or 5 digits is
required to distinguish the results of the two actions
in that regime. The fit of the exponential
decay does not allow for such a high precision.
The behavior of other scaling quantities, like $\Delta E_{3}/
\Delta E_{1}$ etc., is even worse, i.e. the regime of successful
first order perturbation theory is even smaller.
\footnote{As a general trend, the perturbation series gets worse
if higher gaps are involved.}

Actually this construction could be carried on to $O(\tila^{2})$,
but this includes couplings over distances 4, involving up to 8
lattice variables. Moreover, continuum perturbation theory
suggests that the $O(\tila^{2})$ perfect action would only help to 
proceed to slightly larger interactions, see Fig. \ref{figE21}.\\

Still the anharmonic oscillator -- and in particular the ratio
of its mass gaps -- may serve as a good testing ground for
quasi-perfect actions, if we proceed to a {\em non-perturbative
improvement} scheme \cite{prep}. Then one expects a progress also for
moderate and large interactions. 
Thus one overcomes the disadvantages of this model listed at the
end of Section 1, although some truncation of the couplings
will be needed.

In particular, one may introduce
an inverse temperature $\beta$ in the expression for the
partition function plus RGT transformation term, and send
$\beta \to \infty$. Then minimization is sufficient for
a multigrid inverse blocking, and in this way one can identify
a {\em classically perfect action}. Using the standard action
on the fine lattice and a fixed coarse configuration, 
we compared the minima for different
blocking factors, and we typically observed a good convergence 
around blocking factor 10. 
\footnote{For a finite blocking factor $n$ one has to use
the modified RGT parameter $\alpha_{n} = \alpha (1-1/n^{2})$.}
As a test, we run the minimizer at small
$\tila$ and reproduced in this way the 4-spin couplings 
of Section 4.
\footnote{Similarly, for the Schwinger model,
the $O(g)$ (truncated) perfect plaquette couplings \cite{StL} could
be reproduce to percent level from the classically perfect action
\cite{Lang} (even though a slightly different RGT was used).}
The 2-spin couplings $b_{i}$ arise from loop
corrections, hence they are quantum effects, which are
not present in the classically perfect action at small $\tila$.

\begin{figure}[hbt]
\begin{center}
\epsfig{file=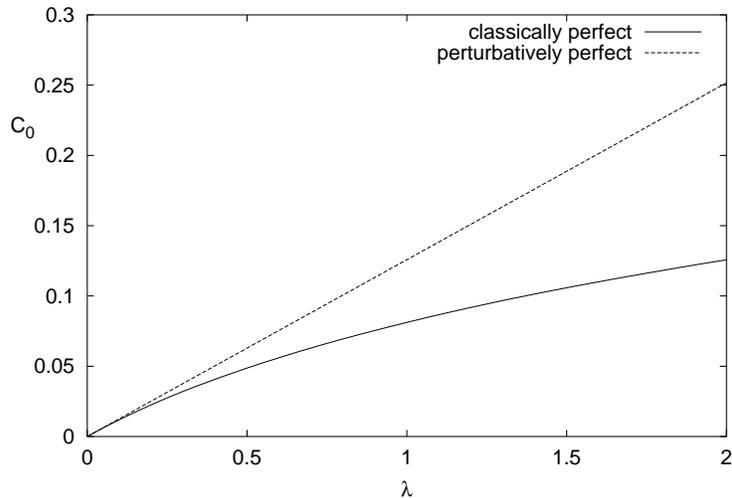,width=10cm,angle=0}
\end{center}
\caption{\it The dominant 4-spin coupling $C_{0}$ at $m=1$ as
a function of the interaction parameter $\lambda $ for the
perturbatively perfect action and for the classically
perfect action.}
\label{figC0m1}
\end{figure}

We can identify the classically perfect couplings over
a wide range of interaction parameters. As an example, Fig.
\ref{figC0m1} compares $C_{0}(\tila )$ for the perturbatively perfect
and for the classically perfect action.
This figure does not imply that the perturbatively perfect
approximation should be rather poor already at 
$\tila \sim 0.1$, as we have to conclude from the simulation
results.

Of course, the model is not asymptotically free, and the 
parameter $\lambda$ is relevant (not just ``weakly relevant'', 
i.e. in leading order marginal), so it does not belong to the class
of models the classically perfect action is designed for. In fact,
from the Table \ref{tabcop} we see that the bilinear contributions
-- that the classical approximation misses -- are important. 
However, it might have a chance to perform well
if we simulate at $\beta >1$, where the quantum corrections
to the perfect action are suppressed.
The possibility that classically perfect actions work to
some extent also for models, which are not asymptotically free,
is conceivable and deserves being tested.
\footnote{One may also think about an extension
to ``semi-classical perfection''.}

Finally, thanks to the
simplicity of the model, one can also perform the full
path integral (numerically) instead, which yields a
(quantum) perfect action. As a test, we also reproduced
roughly some bilinear couplings $b_{i}$ from Section 4 in this way.
There one is restricted to small blocking factors and lattices,
such that a good convergence of the action requires iteration.
Then it is interesting to compare the couplings and their
performance for the classically perfect action, and the
action, which is -- up to numerical errors and truncation --
quantum perfect.
The latter is more promising, but the classically
perfect action is much of interest, because in complicated
field theoretic models, this is what one -- maximally -- has at
hand.

An other important question is the convergence velocity
in the multigrid procedure. In non-Abelian gauge theories, only
very few iteration steps, with a small blocking factor 
(typically 2)
are possible, hence a fast convergence to the FPA is crucial.
We hope that starting from a perturbatively perfect action helps
to accelerate the convergence. Also this can also be tested in
the toy model discussed here.

Furthermore one can use this model to test the ``cycling''
procedure of shifted forward and backward blocking, proposed
by the Boulder group \cite{bold}. It is tractable in higher
dimensions, but not strictly based on the renormalization group.
Hence a toy model analysis of the errors emerging in the
``cycling'' process is of interest.

At last -- as we mentioned in the introduction --
one may speculate that in complicated models the
improvement could be pushed beyond classical perfection, by
performing, say, one full block factor 2 RGT step at finite
$\beta$, starting from a classically perfect action.
This method can also be tested for the anharmonic
oscillator. \\

{\em Acknowledgment} \ \ \
We are indebted to R. Brower, who first suggested this project.
In addition we thank S. Chandrasekharan, S. G\"{u}sken, H. Hoeber, 
Th. Lippert, A. Okopinskaya, G. Ritzenh\"{o}fer, A. Seyfried
and U.-J. Wiese for useful comments.



\end{document}